\documentclass[10pt,twocolumn]{article}

\usepackage{lipsum} 		
\usepackage{blindtext} 		

\usepackage{wrapfig}

\usepackage[superscript, biblabel]{cite}  	

\makeatletter
\renewcommand\@biblabel[1]{#1.}
\makeatother

\usepackage{etoolbox}
\patchcmd{\thebibliography}{\section*{\refname}}{}{}{}

\usepackage[margin=1.0in,hmarginratio=1:1,top=32mm,columnsep=15pt]{geometry}

\usepackage{color}				
\definecolor{dark-gray}{gray}{0.1}	
\usepackage{times}				
\usepackage{microtype} 			
\usepackage[english]{babel} 		

\usepackage[hang, font={footnotesize}, labelfont={bf,up}, textfont={sf,up}]{caption} 
\usepackage{booktabs} 	
\usepackage{mwe}  		

\usepackage{enumitem} 		
\setlist[itemize]{noitemsep} 	

\usepackage{abstract} 
\AtBeginDocument{}  		

\usepackage{titlesec} 			
\renewcommand\thesection{\Roman{section}.} 		  		
\renewcommand\thesubsection{\thesection\Alph{subsection}.} 	
\renewcommand\thesubsubsection{\thesubsection\arabic{subsubsection}.} 
\titleformat{\section}[block]{\normalfont\sffamily\bfseries}{\thesection}{1em}{\MakeUppercase}{} 	
\titleformat{\subsection}[block]{\normalfont\sffamily\bfseries}{\thesubsection}{1em}{}{}  
\titleformat{\subsubsection}[block]{\normalfont\sffamily\bfseries}{\thesubsubsection}{1em}{}{}  
\titlespacing*{\section}{0.0em}{1em}{0.25em}		
\titlespacing*{\subsection}{0.0em}{1em}{0.25em}	
\usepackage{indentfirst}						

\usepackage{fancyhdr} 

\fancypagestyle{plain}{
\fancyhf{}
\lhead{\color{dark-gray}\textit{}} 
\rhead{ {\it Submitted to ANS/NT (2021). LA-UR-21-20901} }
}

\usepackage{titling} 		

\usepackage{hyperref} 	
\hypersetup{
	backref=true,       
    	pagebackref=true,               
    	hyperindex=true,                
    	colorlinks=true,                
    	breaklinks=true,                
    	urlcolor= black,                
    	linkcolor=blue,                
    	bookmarks=true,                 
    	bookmarksopen=false,
    	filecolor=black,
    	citecolor=blue
}

\pretitle{\begin{center}\large\bfseries} 	
\posttitle{\end{center}} 				
\title{\vspace{-0.3in} \sffamily{Criticality Experiments with Fast 25 and 49 Metal and Hydride Systems During the Manhattan Project} }	
\author{%
\normalsize Jesson Hutchinson\thanks{corresponding author: jesson@lanl.gov}, Jennifer Alwin, Alexander McSpaden, William Myers, Michael Rising, and Rene Sanchez \\[-0.5ex] 
\normalsize Los Alamos National Laboratory \\[-0.5ex] 
\normalsize Los Alamos, NM 87545
}
\date{ } 
\renewcommand{\maketitlehookd}{%
\begin{abstract}
\vspace*{-0.5in}
\noindent {\normalfont\textbf{Abstract}\textemdash}
\noindent 
Criticality experiments with $^{235}$U (metal and hydride) and $^{239}$Pu (metal) were performed during the Manhattan Project. Results from these experiments provided necessary information for the success of the Manhattan Project. These experiments have been previously described in compilations made after the Manhattan Project~\cite{HisExp_1,HisExp_2,HisExp_3}, but those works are either lacking in technical details or are not publicly available. This work aims to provide detailed information while showcasing the enduring impact of these experiments 75 years after they were performed. Furthermore, we use modern computational methods embodied in the MCNP6 code and ENDF data to analyze and interpret these historic measurements. The world's first four criticality accidents are also discussed, as lessons learned from these helped shape the field of criticality experiments.
\end{abstract}
}

\begin{document}

\maketitle	


\section{Introduction}
\label{sec:Intro}
When the Los Alamos laboratory was formed in April 1943, Fermi had already achieved a critical configuration in CP-1~\cite{HisExp_4},  but many questions remained related to criticality. As stated by Robert Serber in the Los Alamos Primer~\cite{HisExp_5},  “these values of critical masses are still quite uncertain, particularly those for 49\footnote{During the Manhattan Project, a notation in which the last integer of the Z number and A number are combined to refer to a nuclide. For example 25, 28, and 49 refer to $^{235}$U, $^{238}$U, and $^{239}$Pu respectively. This notation is retained throughout this work. $^{240}$Pu was called both 40 and 410, but this work will refer to it only as 40.}. To improve our estimates requires a better knowledge of the properties of bomb materials and tamper: neutron multiplication number, elastic and inelastic cross-sections, overall experiments on tamper materials. Finally, however, when materials are available, the critical masses will have to be determined by actual test.” These types of measurements were then performed during 1944 and 1945, as described below and in our companion paper on the Water Boiler and Dragon~\cite{HisExp_6}. The results of these criticality experiments were so important that Oppenheimer stated to “notify him at once if any serious change in the estimated critical mass should be indicated by the experiments”~\cite{HisExp_9}.

The focus of this work is on two specific experiments: one using metal $^{235}$U and $^{239}$Pu in spherical geometry and one utilizing cubes of $^{235}$U in hydride form. These were the first experiments performed in the world with large (\textgreater100 g) quantities of pure nuclear material. This material did not start arriving until late 1944 and experiments were performed as new material kept arriving in Los Alamos. Many of these experiments were subcritical, as there was not enough material available yet to obtain criticality. Both experiments included bare (no reflector) measurements as well as measurements with various reflectors. All of the bare experiments, however, were subcritical and no bare critical assemblies were constructed until 1951 for 25~\cite{HisExp_10,HisExp_11}, and 1954 for 49~\cite{HisExp_12,HisExp_13}. These experiments helped answer key questions necessary for the Manhattan Project to be successful. 

This paper includes information on the background and theory associated with criticality experiments, early nuclear data experiments, and a description of the metal and hydride criticality experiments. Information of early criticality accidents is also presented. Lessons learned from these accidents are as important as the experiments themselves, since they resulted in the establishment of a critical experiment capability. Last, the continued impact of these experiments and accidents is discussed. References in this work include many source documents that may not be easily accessible by the public.

\section{Background and Theory}
\label{sec:Background}
During the Manhattan Project, multiplication ($M$) was defined as 
\begin{equation}
M=1+\frac{N_f}{N_0}(\nu-1-\alpha)
\label{eq:M}
\end{equation}
where $N_f$ is the number of fissions produced by $N_0$ source neutrons, $\nu$ is the number of neutrons produced per fission, and $\alpha$ is the ratio of capture to fission~\cite{HisExp_14,HisExp_15}. Note that today this is often referred to as “total multiplication” and relates to the effective multiplication factor via
\begin{equation}
M=\frac{1}{1-k_{eff}}
\label{eq:keff}
\end{equation}
Fermi established the approach-to-critical method used at CP-1. As they were building CP-1, Fermi and others were taking measurements of the neutron density in center of the pile with indium foils. The activity of these foils was measured and recorded as a function of the number of natural uranium layers as they were building CP-1. The inverse of the geometry corrected activity of the indium foils was plotted as a function of natural uranium layers. The plot clearly predicted the layer (amount of natural uranium) at which CP-1 was going to go critical~\cite{HisExp_4}. A similar method is used to this day for assembling fissile material, such as the experiments performed at the National Criticality Experiments Research Center (NCERC) in Nevada~\cite{NCERChayes2016ANTPC}. For this method, relative multiplication is approximated by measuring count rates of two systems, often using external detectors, via
\begin{equation}
\frac{C}{C_0}=\frac{\epsilon S M \Omega}{\epsilon_0 S_0 M_0 \Omega_0}
\label{eq:approach}
\end{equation}
where $C$ is the detector count rate, $S$ is the source emission rate, $\Omega$ is the detector solid angle, and $\epsilon$ is the detector intrinsic efficiency. Here the 0 subscript refers refers to an initial configuration, and the multiplication is relative to the initial configuration as shown in the equation. This initial configuration may be a system with no material (other than a source), which will have $M=1$. Ideally, for properly designed experiments, the following terms will not change as the configuration changes (through addition of fuel or reflector material for example): $S$, $\Omega$, and $\epsilon$. If those terms are constant, then Equation~\ref{eq:approach} reduces to
\begin{equation}
\frac{C}{C_0}=\frac{\epsilon S M \Omega}{\epsilon_0 S_0 M_0 \Omega_0}=\frac{M}{M_0}=M
\label{eq:approach_simple}
\end{equation}
and the multiplication can be estimated. If the multiplication of two configurations are measured (often one of the configurations may have no fissionable material and therefore $M = 1$), then the reciprocal of the multiplication ($1/M$) can be plotted and the critical configuration (often critical mass, but it could be critical radius or reflector thickness) can be estimated.

Once the critical mass was determined, it was often convenient to estimate the equivalent mass for different densities and enrichments. During the Manhattan Project, the following empirical estimates were given~\cite{HisExp_16,HisExp_17}:
\begin{equation}
critical~mass \sim \rho^{-1.4} C^{-1.8}
\label{eq:crit_mass_correction}
\end{equation}
\begin{equation}
critical~radius \sim \rho^{-0.8} C^{-0.6}
\label{eq:crit_radius_correction}
\end{equation}
where $\rho$ is the density and $C$ is the enrichment. It is unclear why the exponent for $\rho$ is $-$1.4 and not $-$2; today (and for at least the last 40 years), it is common knowledge that critical mass varies inversely with the square of the density. That being said, much less information was available, as these were the first measurements ever performed of this type. This is investigated in Section~\ref{sec:spheres}. Additional corrections were also made to account for other parameters (such as geometry effects or reflector density).

The two experiments described in this work occurred in 1944 and 1945 and evolved as more material was made available. These experiments took place after the Water Boiler~\cite{HisExp_6} had started up (since that experiment required less nuclear material). These experiments utilized much of the same equipment and personnel as the Water Boiler experiments.

A large amount of coordination was needed to achieve these criticality experiments. Due to the importance and the complexity of this coordination, Oppenheimer and the technical board generally advised on the prioritization and scheduling of these experiments (which took place during meetings in Oppenheimer’s office, often at night)\footnote{See “Minutes of Technical and Scheduling Conference” meeting notes by Samuel Allison from 1944-1945.}. Coordination was required in regards to receipt of material from various sites, metallurgy, design of experiments, execution of experiments, measurement equipment, and theoretical calculations (in addition to other considerations). Often a prioritized list of experiments was established, and sometimes not all the experiments were performed, as other uses of the same nuclear material were deemed to be of higher priority.

\section{Nuclear Data}
\label{sec:ND}
At the start of the Manhattan Project, the collective knowledge of nuclear data was limited. As an example, the number of neutrons produced per fission ($\nu$) was unknown for fission in 49 and for fast fission in 25~\cite{HisExp_2,HisExp_5,HisExp_18}. It was understood from the beginning of the Manhattan Project that understanding of the properties of fission and radiation interaction in matter was required for success. These measurements started in 1943 and were often compared to hand calculations. Many of the measurements were compared to measurements taking place at other laboratories, such as the Metallurgical Laboratory in Chicago. The majority of these measurements were conducted with small quantities of material (often milligram samples), and the estimates would be refined based on the material used and the techniques used for data analysis. It was very important to have repeatable measurements as the nuclear data was so uncertain and new at that time.

\begin{figure*}[htb!]
	\centering\includegraphics{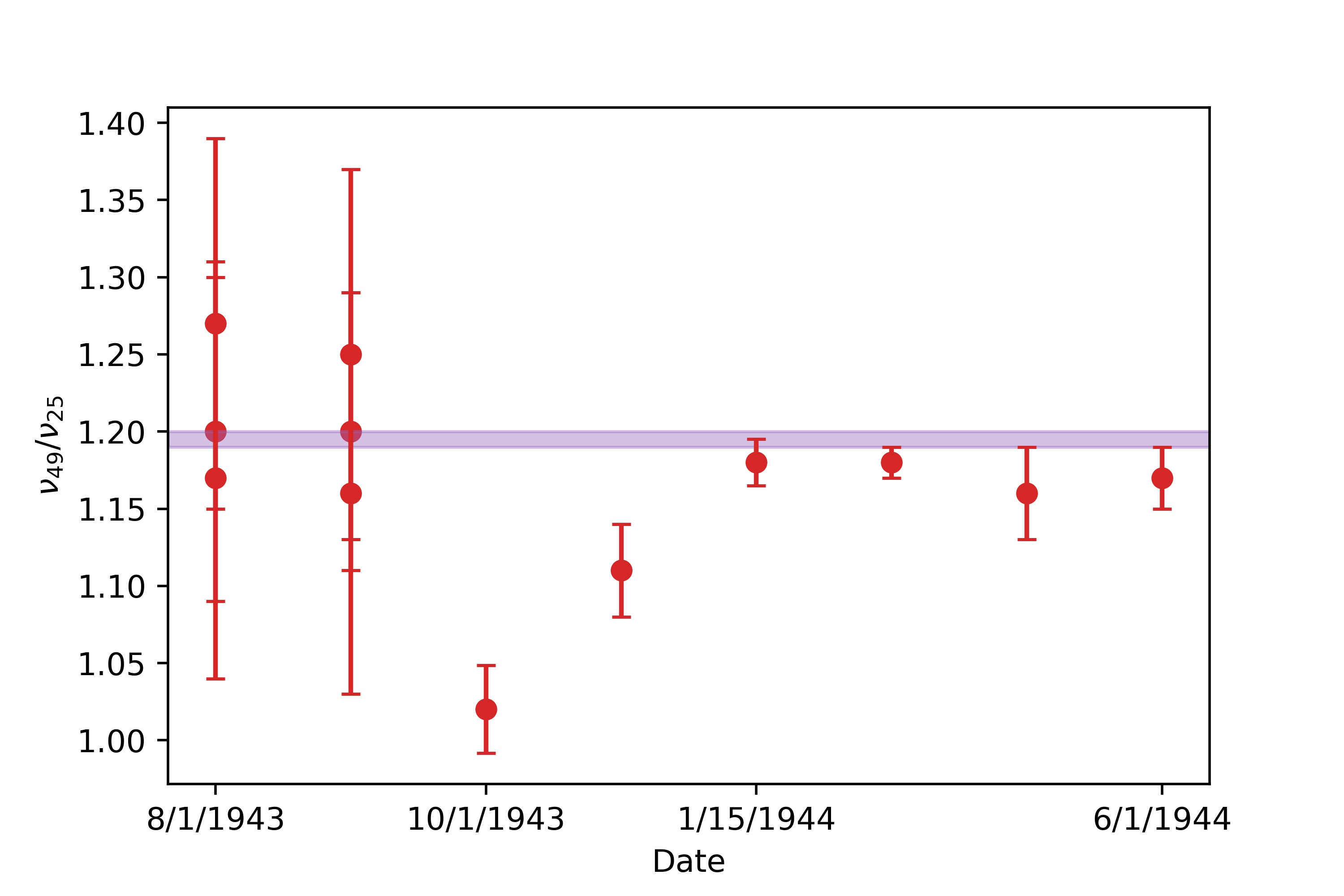}
	\caption{Ratio of the number of neutrons emitted from fission ($\nu$) in 49 to 25. Data points were all measured during the Manhattan Project. The shaded region is 1.19–1.20, the ratio from ENDF/B-VIII.0~\cite{ENDF8}  (assuming 1–2 MeV incident neutron energy).} 
	\label{Fig_F1_nu40_nu25}
\end{figure*}

The number of neutrons produced per fission ($\nu$) was of extreme importance (and remains important to this day) as it will affect criticality more than any other nuclear data parameter. Many measurements of $\nu$ were performed during the Manhattan Project. A subset of these include measurements of 23~\cite{HisExp_19}, 25~\cite{HisExp_20}, 40 spontaneous fission~\cite{HisExp_21}, and 49~\cite{HisExp_19,HisExp_20,HisExp_22,HisExp_23}. Often 25 was (and still is) used as a standard for relative comparisons. Figure~\ref{Fig_F1_nu40_nu25} shows various measurements of $\nu_{49}/\nu_{25}$ performed during the Manhattan Project~\cite{HisExp_22,HisExp_23,HisExp_24,HisExp_25,HisExp_26,HisExp_27,HisExp_28} and compares them against ENDF/B-VIII.0~\cite{ENDF8}, the latest US nuclear data library release. It can be seen that the last four data points in Figure~\ref{Fig_F1_nu40_nu25} are in good agreement with the current ENDF/B-VIII.0 data.

Fission cross-sections were (and still are) also of great importance. During the Manhattan Project, the personnel knew of resonances and were able to estimate some of them~\cite{HisExp_20}. It was known that some resonances could not be resolved at that time~\cite{HisExp_29}.  The majority of the fission cross-section measurements were on induced fission in 25~\cite{HisExp_14,HisExp_20,HisExp_21,HisExp_28,HisExp_31,HisExp_32,HisExp_33,HisExp_34,HisExp_35,HisExp_36,HisExp_37,HisExp_38,HisExp_43}. In addition, measurements were performed for 23~\cite{HisExp_19,HisExp_21}, 28~\cite{HisExp_14,HisExp_20,HisExp_32,HisExp_36,HisExp_39}, 37~\cite{HisExp_32,HisExp_39}, and 49~\cite{HisExp_28,HisExp_31,HisExp_32} fission. 25 was again used as a standard, and the ratio of induced fission in 49 to 25 was often reported~\cite{HisExp_20,HisExp_23,HisExp_41}. 

Other reaction cross-sections investigated included capture and scattering. Capture was investigated for many elements, including gold~\cite{HisExp_25,HisExp_43,HisExp_44,HisExp_45,HisExp_46}   (which is a standard to this day), 25, and 28~\cite{HisExp_33,HisExp_47,HisExp_48}. One parameter often reported is $\alpha$, which is the ratio of capture to fission. This was measured for 25~\cite{HisExp_49} and 49~\cite{HisExp_28}. Similarly, scattering was measured for 25~\cite{HisExp_33,HisExp_50,HisExp_51} and 28~\cite{HisExp_52},  among other nuclides and elements.

Before material was available to investigate criticality, semi-integral measurements were already estimating the effectiveness of different reflector materials~\cite{HisExp_36}. The materials that were the most effective at reflecting neutrons were studied often (theory was developed to understand this effectiveness)~\cite{HisExp_53}.  These included Tuballoy (Tu, natural uranium)~\cite{HisExp_33,HisExp_35,HisExp_45,HisExp_54,HisExp_55,HisExp_56}, beryllium (Be)~\cite{HisExp_57}, tungsten (W)~\cite{HisExp_33,HisExp_35,HisExp_45,HisExp_54,HisExp_55,HisExp_56}, tantalum (Ta)~\cite{HisExp_33,HisExp_45,HisExp_54,HisExp_56}, and lead (Pb)~\cite{HisExp_33,HisExp_35,HisExp_45,HisExp_54,HisExp_55,HisExp_56}. Many of these measurements were types of transmission measurements, which are commonly performed to this day.

Spontaneous fission measurements, led by Emilio Segrè, were also performed throughout the Manhattan Project. The most notable measurements were of 40, which confirmed suspicions by Seaborg, Fermi, and others that 40 spontaneous fission was very high~\cite{HisExp_3,HisExp_18,HisExp_59}. These results were the main driver for the reorganization of Los Alamos in 1944~\cite{HisExp_3,HisExp_18,HisExp_59}. In addition to the 40 measurements, additional measurements were performed on 25~\cite{HisExp_31,HisExp_60}, 28~\cite{HisExp_43,HisExp_60}, 48~\cite{HisExp_21}, and 49~\cite{HisExp_35,HisExp_36,HisExp_45}. Comparison between 49 produced in Berkeley (likely 99+\% 49) and Clinton Engineer Works (which would have higher 40 content) is what led to the initial suspicion that 40 spontaneous fission may be very high~\cite{HisExp_35,HisExp_36}. One important note is that these were precision measurements, as any measurements on nuclides that have low spontaneous fission emission require both very long count times and very low, well-characterized background. Additional information on nuclear data from the Manhattan Project is given in a different work in this same issue~\cite{HisExp_74}.

\section{25 and 49 metal spherical experiments}
\label{sec:spheres}
Measurements were performed to estimate the multiplication of 25 and 49 metal spheres. The 25 was in the form of $\beta$-stage (72–81\% enrichment 25)~\cite{HisExp_16} material. These measurements continuously took place from late-1944 until mid-1945 and were performed by R division (led by Robert Wilson). 

\begin{figure*}[htb!]
	\centering\includegraphics[width=6in]{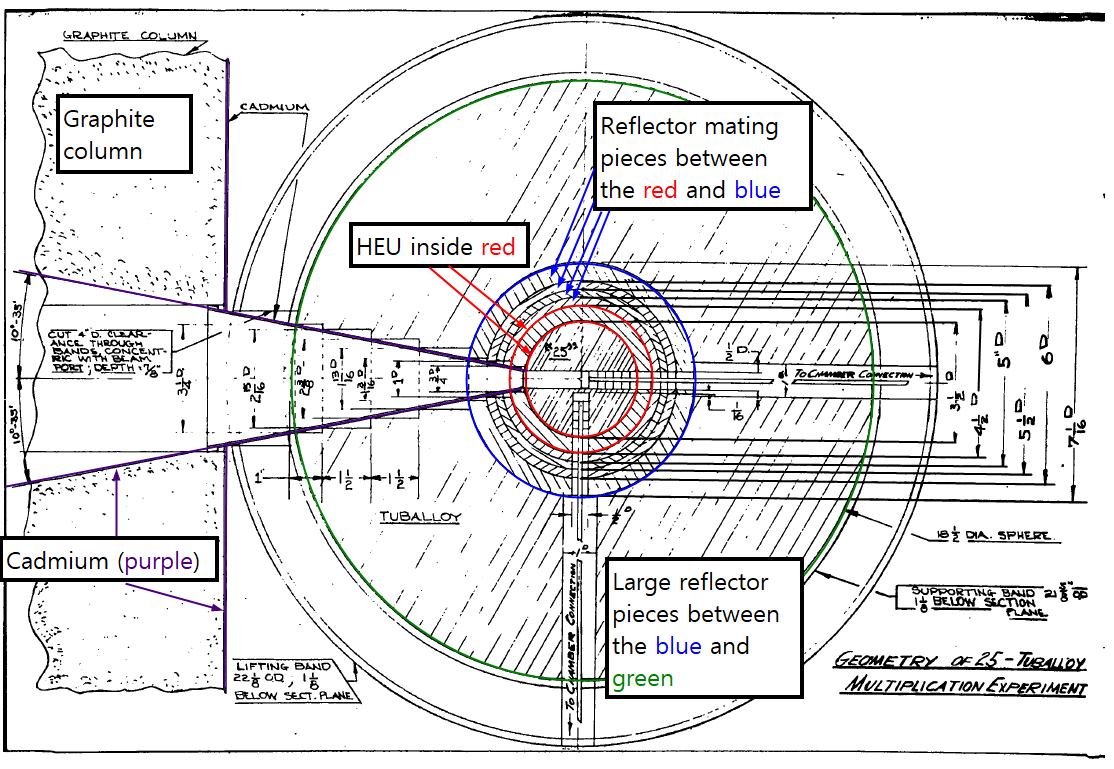}
	\caption{Setup of the fission chamber measurements (which inferred system multiplication) for the 25 metal sphere experiment reflected by Tu with the Water Boiler as the neutron source (from~\cite{HisExp_16}).} 
	\label{Fig_F2_sphere_setup}
\end{figure*}

As additional metal was made available, new measurements were performed, as shown in Table~\ref{Tab:1_sphere_history}. These parts were referred to by their outer diameter (OD), which is also retained here. The initial 25 material had a 0.314 inch ID and 1.5 inch OD, which had a mass of 525 grams. As seen in Table~\ref{Tab:1_sphere_history}, more material was used up to an OD of 4.5 inches, which had a corresponding mass of 14–15 kg (not reported, but calculated using reported dimensions and a density of 18.4 g/cm$^3$). A 5.0 inch OD metal sphere was planned (which would have been the last 25 metal sphere)~\cite{HisExp_9,HisExp_61}. This material was originally scheduled to be available on March 15$\textendash$20, 1945. Due to the 4.5 inch OD results as well as the urgent overall schedule of the Manhattan Project, it was decided that it was not necessary to make the 5.0 inch OD sphere~\cite{HisExp_62}.  Some of these hemispheres nested together as shown in Figure~\ref{Fig_F2_sphere_setup}. 

\begin{table*}[htb!]
\caption{Material used in 25 and 49 metal sphere experiments.}
\centering{}%
\begin{tabular}{|c|c|c|}
\hline 
Date of Report & Material Type and OD (inches) & Reflectors\tabularnewline
\hline 
\hline 
10/1/1944~\cite{HisExp_33} & 25: 1.5" & None\tabularnewline
\hline 
11/1/1944~\cite{HisExp_63} & 25: 1.5", 2" & None\tabularnewline
\hline 
12/18/1944~\cite{HisExp_67} & 25: 1.5", 2" & None\tabularnewline
\hline 
12/20/1944~\cite{HisExp_60} & 25: 1.5", 2", 2.5" & None\tabularnewline
\hline 
3/1/1945~\cite{HisExp_21} & 25: 3.5" & None, Tu, WC\tabularnewline
\hline 
4/11/1945~\cite{HisExp_14} & 25: 1.5", 2", 2.5". 49: 0.9" & None\tabularnewline
\hline 
4/14/1945~\cite{HisExp_61} & 25: 1.5", 2", 2.5", 3.5", 4.5" & None, Tu, WC\tabularnewline
\hline 
10/6/1945~\cite{HisExp_16} & 25: 3.5", 4.5" & Tu, WC, WC+Fe\tabularnewline
\hline 
10/30/1945~\cite{HisExp_17} & 25: 3.5", 4.5" & Tu, WC\tabularnewline
\hline 
\end{tabular}\label{Tab:1_sphere_history}
\end{table*}

Tuballoy and tungsten carbide (WC) reflection were provided by some experiments. The Tu consisted of two large hemishells with 7.0625 inch ID and 18.125 inch OD (approximately 900 kg); smaller hemishells were used to mate these reflectors to the 3.5 inch or 4.5 inch OD 25 spheres as shown in Figure~\ref{Fig_F2_sphere_setup}. The WC reflectors were constructed from 2.125 x 2.125 x 4.25 inch blocks. It is assumed that these blocks were arranged to form a pseudo-sphere.  

When performing criticality experiments with HEU and pure 49, it is necessary to have a source of neutrons in order to obtain statistically significant results in reasonable measurement times. Various neutron sources were used for these experiments. Mock fission sources which utilized a mixture of Po (an $\alpha$ emitter) with either B~\cite{HisExp_64},  BF$_3$~\cite{HisExp_65},  NaBF$_4$~\cite{HisExp_66},  or a mixture of NaBF$_4$ and BeF$_2$~\cite{HisExp_19,HisExp_66} were used~\cite{HisExp_14,HisExp_33,HisExp_60,HisExp_61,HisExp_63}. The mock fission sources improved during the Manhattan Project, and different target materials were chosen to provide a better match of the prompt fission neutron spectrum. Instead of mock fission sources, some experiments utilized the Water Boiler~\cite{HisExp_16,HisExp_17,HisExp_60}. For these experiments, the Water Boiler was operated up to 4.5~kW~\cite{HisExp_17} and a graphite column was utilized with a cadmium (Cd) cone to ensure that only direct neutrons (those that have not undergone scattering) from fission in the Water Boiler reached the spheres, as shown in purple in Figure~\ref{Fig_F2_sphere_setup}.

The criticality experiments on the metal spheres were measured using multiple means. Several experiments used the long counter~\cite{HisExp_14,HisExp_33,HisExp_61,HisExp_63}, which would be external to the spheres. The long counter was designed to have uniform efficiency over a large range of neutron energies and had BF$_3$ inside paraffin~\cite{HisExp_68,HisExP_69}. Other experiments utilized 25 and 28 fission chambers~\cite{HisExp_16,HisExp_60,HisExp_67} and cellophane foils~\cite{HisExp_17} inside the spheres. Geiger-Müller counters were used to measure fission fragments from the cellophane foils. Generally, the long counter was used for the experiments with the mock fission source, and the fission chambers and foils were utilized for the Water Boiler experiments.

\begin{figure*}[htb!]
	\centering\includegraphics[width=5in]{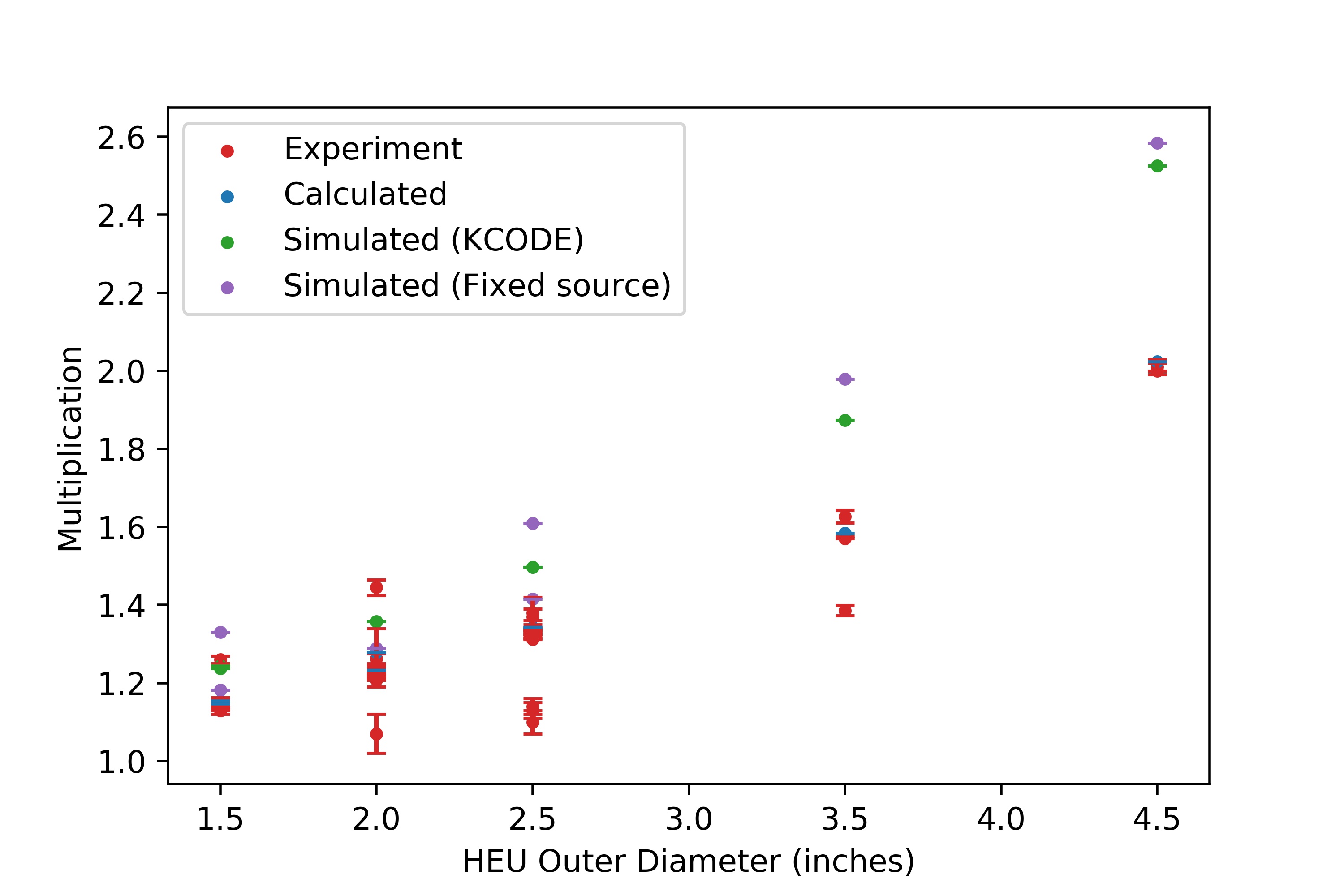}
	\caption{Multiplication (measured and calculated) for the bare 25 sphere experiments. Results are from~\cite{HisExp_14,HisExp_21,HisExp_60,HisExp_61,HisExp_63,HisExp_67} (all reported during the Manhattan Project).} 
	\label{Fig_F3_sphereM_bare}
\end{figure*}

\begin{figure*}[htb!]
	\centering\includegraphics[width=5in]{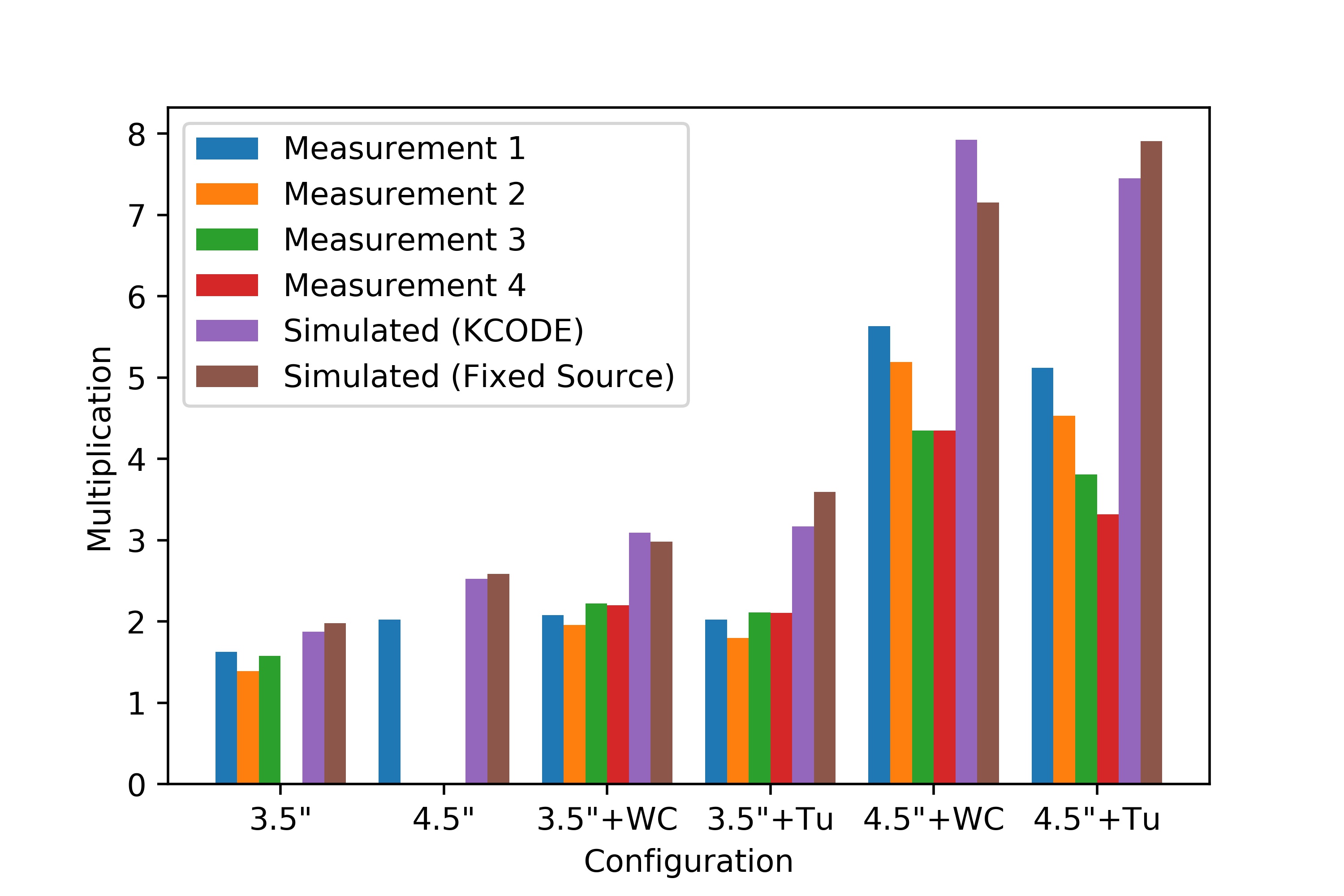}
	\caption{Multiplication (measured) for the bare and reflected 25 sphere experiments. Results are from~\cite{HisExp_16,HisExp_17,HisExp_21,HisExp_61} (all reported during the Manhattan Project).} 
	\label{Fig_F4_sphereM_ref}
\end{figure*}

Results from the bare experiments are given in Figure~\ref{Fig_F3_sphereM_bare} and for the reflected experiments in Figure~\ref{Fig_F4_sphereM_ref}. The measured results were obtained using Equation~\ref{eq:approach_simple}; since the initial count was generally taken without any material present (just a source), the "$M$"  given in the equation is an estimate of the absolute multiplication. The bare results are compared with calculated results (from the same references), but no calculated results were provided for the reflected configurations. It can be seen in these figures that there was a fairly wide range of results. However, it should be noted that even today, when at low multiplication, the uncertainties are often quite large. The uncertainties shown in Figure~\ref{Fig_F3_sphereM_bare} only include statistical uncertainties, and it is known that these measurements would have much larger systematic uncertainties, which were not estimated. Given that multiple detection systems and analysis methods were used, the spread in results shown is not surprising. The calculated results were determined using neutron diffusion theory~\cite{HisExp_15,HisExp_70}.

Figures~\ref{Fig_F2_sphere_setup} and~\ref{Fig_F3_sphereM_bare} include results using MCNP6.2\textregistered\footnote{MCNP\textregistered ~and Monte Carlo N-Particle\textregistered ~are registered trademarks owned by Triad National Security, LLC, manager and operator of Los Alamos National Laboratory. Any third party use of such registered marks should be properly attributed to Triad National Security, LLC, including the use of the designation as appropriate. For the purposes of visual clarity, the registered trademark symbol is assumed for all references to MCNP within the remainder of this paper.} with ENDF/B-VIII.0 data. These include both criticality eigenvalue (KCODE) and fixed source simulation results. It can be seen that the comparison to measurements is not particularly close. That being said, it should be stressed that the documentation on these experiments is not adequate to create a detailed model. So there certainly could be a bias in the models due to this lack of documentation.

Many corrections or extrapolations were applied to the results. These included corrections for the small central cavity (to allow for fission chambers or foils)~\cite{HisExp_71}, corrections to higher (nominal) densities using Equation~\ref{eq:crit_mass_correction}~\cite{HisExp_16}, corrections to higher enrichment using Equation~\ref{eq:crit_mass_correction}~\cite{HisExp_16}, reflector thickness~\cite{HisExp_16}, and reflector impurities~\cite{HisExp_72}. 

\begin{figure*}[htb!]
	\centering\includegraphics[width=5.2in]{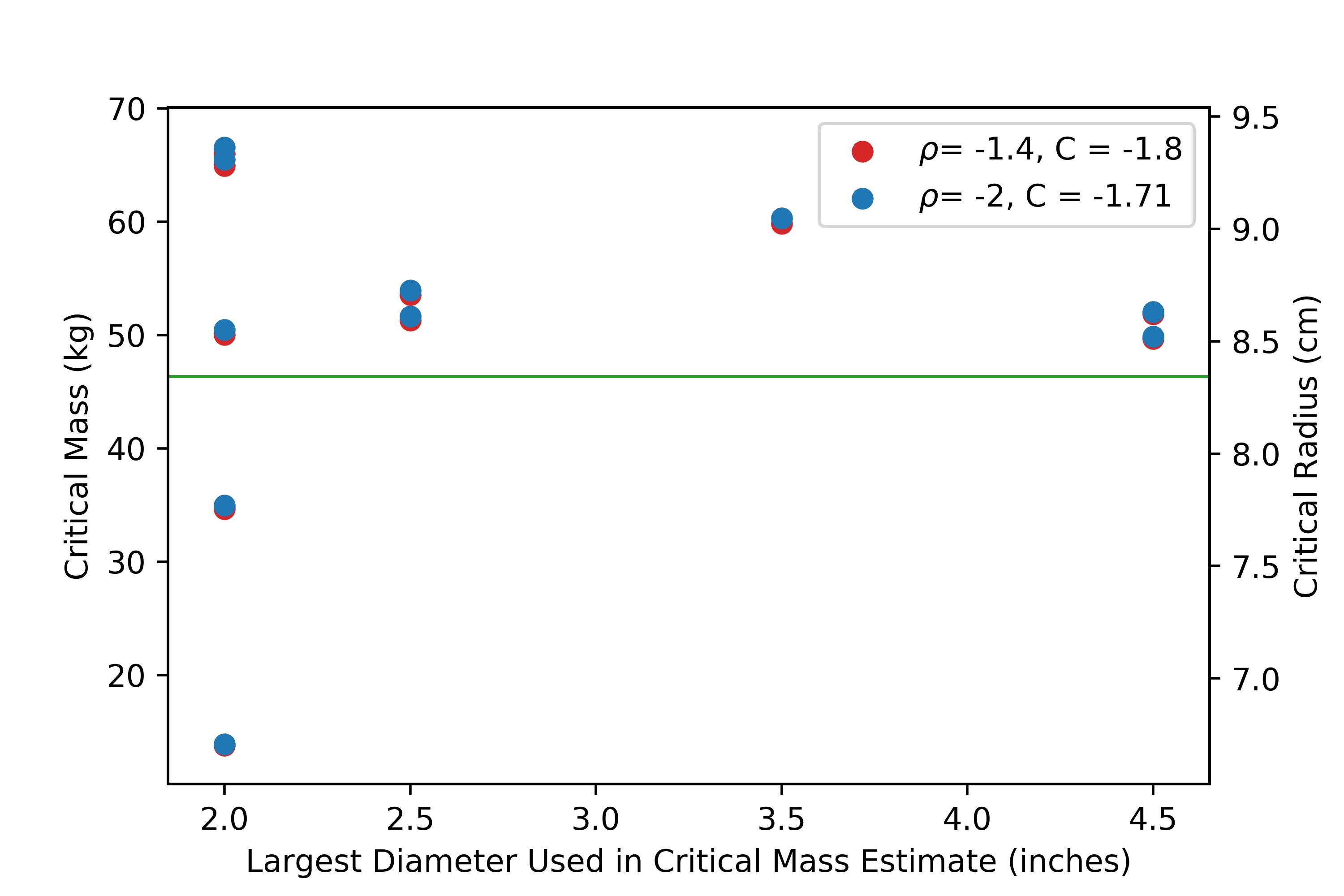}
	\caption{Critical mass and radius estimates for bare 25 (100\% enrichment). These estimates utilize multiplication results from references~\cite{HisExp_14,HisExp_60,HisExp_61,HisExp_63}.} 
	\label{Fig_F5_crit_mass_and_rad}
\end{figure*}

Ironically, no bare critical mass estimates were given in the references, even though the measurements can be (and were) used to estimate the bare critical mass. Here we will describe how to perform this estimate with the original data. In order to estimate the bare critical mass, the inverse of the multiplication values in Figure~\ref{Fig_F3_sphereM_bare} were first used to infer a critical mass with the density and enrichment associated with the measured HEU. Then Equation~\ref{eq:crit_mass_correction} was used to estimate the critical mass of pure 25 at nominal density (19 g/cm$^3$). Note that no correction was made for the small central hole; this, however, would have a very small effect on the results (especially those that use larger hemishells). It should also be noted that some of the references are inconsistent in regards to enrichment and density of the 25 spheres; this is particularly important for enrichment, as the critical mass changes about 1 kg per 1\% change in enrichment. The exponent values in Equation~\ref{eq:crit_mass_correction} are different from those used “today”: $-$2 (not $-$1.4) is used for density correction~\cite{HisExp_76} and $-$1.71 (not $-$1.8) was given in the 1950s after more material was available~\cite{HisExp_73}.  In the end, these differences in corrections do not matter very much, as shown in Figure~\ref{Fig_F5_crit_mass_and_rad}, which shows the bare 25 critical mass estimates. The critical radius is found using the same data with Equation~\ref{eq:crit_radius_correction} to correct for an ideal sphere of 25 and is shown on the right y-axis. 

\begin{figure*}[htb!]
	\centering\includegraphics[width=6in]{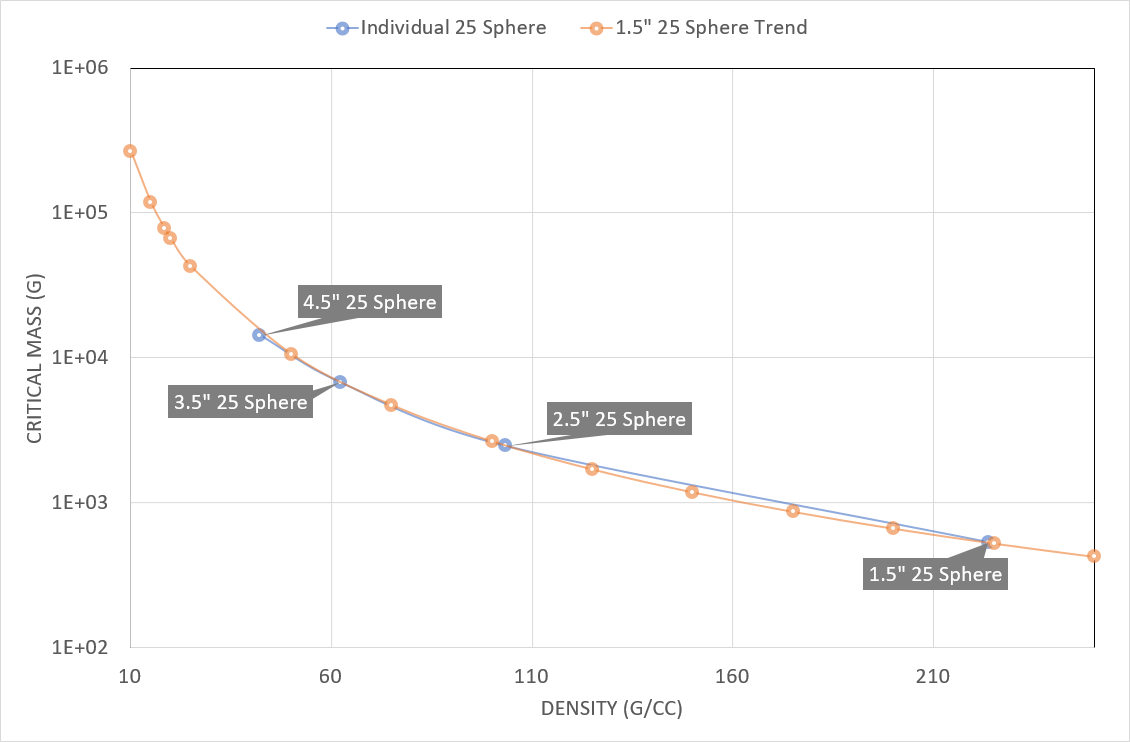}
	\caption{MCNP simulations of critical mass versus density using the 25 metal sphere specifications.} 
	\label{Fig_N1_crit_mass_vs_density}
\end{figure*}

In order to confirm that the density exponent should indeed be $-$2, a series of MCNP simulations were performed as shown in Figure~\ref{Fig_N1_crit_mass_vs_density}. First the enrichment information of the 1.5 inch OD sphere was used, along with a chosen density, and the radius was varied to find a critical mass. In addition, each of the 25 sphere models (1.5, 2.5, 3.5, and 4.5 inch OD) was used (keeping the geometry constant this time), and the density was varied (shown in blue). This resulted in an exponent of $-$1.979, thus confirming the expected result of $-$2 (not $-$1.4).  

It can be seen in Figure~\ref{Fig_F5_crit_mass_and_rad} that the critical mass estimates using these data vary widely. This is not surprising, however, given the spread in the results in multiplication shown in Figure~\ref{Fig_F3_sphereM_bare}. Since the predicted critical is related to the slope of inverse multiplication, a large spread in multiplication will result in a large spread of critical mass. It is known, however, that as the multiplication increases, the accuracy of the critical prediction improves. For this reason, the results in Figure~\ref{Fig_F5_crit_mass_and_rad} were plotted as a function of the largest hemishells used for the critical mass estimate. So the points on the right side of the graph are expected to be more accurate (and have less spread) than the points on the left side. It can be seen that this is indeed the case when compared against today’s ENDF/B-VIII.0 result of 46.36 kg~\cite{ENDF8}. Enough information to provide accurate uncertainty estimates is not available (especially in regards to enrichment). Using the two data points on the far right of Figure~\ref{Fig_F5_crit_mass_and_rad}, we would give a best estimate of the critical mass from these experiments as 50.8 $\pm$ 2.5 kg. These critical mass estimates should also be compared against calculations of 25 critical mass, given in this issue~\cite{HisExp_74}. 

The same approach can also be applied to the Lady Godiva experiment performed at Los Alamos in the 1950s~\cite{HisExp_10}. Unlike the deeply subcritical spherical experiments discussed in this section, Lady Godiva was a bare sphere of HEU that was critical. The Lady Godiva experiment had a mass of 52.42 kg, a density of 18.74 g/cm$^3$, and an enrichment of 93.71 wt.\% 25~\cite{HisExp_75}.  Using these specifications with Equation 5, we calculate a critical mass of 45.7 kg (using the exponent values in Equation 5). Using the newer exponent values ($-$2 for density and $-$1.71 for enrichment as described above), we calculate a similar value of 45.6 kg. Both of these are very close to the ENDF/B-VIII.0 result of 46.36 kg, which should not be surprising since ENDF data has typically been calibrated to accurately match the Lady Godiva results.

The measurements given in Figure~\ref{Fig_F4_sphereM_ref} were used in Reference~\cite{HisExp_17} to estimate the critical mass for 25 reflected by WC and Tu: these estimates were 13.8 kg and 15.8 kg, respectively. Commonly reported values for these~\cite{HisExp_76}  include 16.0 kg for WC and 16.1 kg for Tu. Using MCNP6.2 with ENDF/B-VIII.0 gives 14.96 kg for the critical mass of 100\% 25 at 19 g/cm$^3$ surrounded by infinite WC at 15 g/cm$^3$.

Due to the high toxicity of 49, it was desirable to reduce the number of refabrications and the number of individual pieces~\cite{HisExp_1}. A 49 sphere with 0.9 inch OD was available around March$\textendash$April 1945. Information on fabrication of this 49 is given in a different work in this issue~\cite{HisExp_77}.  This resulted in a reported multiplication of 1.197 $\pm$ 0.004. This measurement is significant, as it is the first recorded measurement of multiplication in 49. Prior to this, only small quantities of 49 were available. The exact mass of the sphere is unknown, but it was likely around 100 g.

The metal experiments were utilized to estimate the quantity $\nu-1-\alpha$. This resulted in 25 values shown in Table~\ref{Tab:2_nu-1-alpha}, which also compares them to modern results. Only two results for 49 were given during the Manhattan Project, which were 2.14 $\pm$ 0.1614 and 2.02~\cite{HisExp_1,HisExp_78}; these compare well to a value calculated from modern nuclear data of approximately 2.08 $\pm$ 0.10~\cite{HisExp_42}. 

\begin{table}[htb!]
\caption{Material used in 25 and 49 metal sphere experiments.}
\centering{}%
\begin{tabular}{|c|c|}
\hline
$\nu-1-\alpha$ & Reference\tabularnewline
\hline
\hline
1.25 $\pm$ 0.01 & LAMS-163\tabularnewline
\hline 
1.13 $\pm$ 0.01	& LAMS-163\tabularnewline
\hline 
1.44 $\pm$ 0.02 & LAMS-163\tabularnewline
\hline 
1.21 $\pm$ 0.04 & LAMS-163\tabularnewline
\hline 
1.08 $\pm$ 0.008 & LAMS-175\tabularnewline
\hline 
1.51 $\pm$ 0.05 & LAMS-227\tabularnewline
\hline 
1.52 $\pm$ 0.08 & LAMS-227\tabularnewline
\hline 
1.51 & LA-140 A\tabularnewline
\hline 
1.57 $\pm$ 0.03 & LA-140 A\tabularnewline
\hline 
1.54 & LA-464\tabularnewline
\hline 
1.52 & LA-1033\tabularnewline
\hline 
1.54 $\pm$ 0.08 & LA-1033\tabularnewline
\hline 
1.67 & LA-1033\tabularnewline
\hline 
1.59 $\pm$ 0.08 & LA-1033\tabularnewline
\hline 
1.52 $\pm$ 0.09 & ENDF/B-VIII.0\tabularnewline
\hline 
\end{tabular}\label{Tab:2_nu-1-alpha}
\end{table}

\section{Hydride assembly}
\label{sec:hydride}
Criticality experiments with solid uranium hydride were performed in 1944–1945 by group G-1 (led by Otto Frisch underneath division leader Robert Bacher). The use of uranium hydride was logical to the fact that less material would be required due to moderation created by the presence of hydrogen. 

The fissile material consisted of UH$_3$ and Styrex (CH$_{1.75}$) in the form of 0.5 inch or 1.0 inch cubes. Initially, enough material was not available to be critical with UH$_{10}$ using a BeO reflector. For this reason, initial experiments had a higher hydrogen content (UH$_{80}$, then UH$_{40}$, UH$_{25}$, UH$_{15}$, until it was possible to go critical with UH$_{10}$). In September 1944, 0.25 kg could be produced in one day but by December this increased to 0.5$\textendash$1 kg~\cite{HisExp_80}. The information presented in this section largely comes from a single report, which was not documented until after the Manhattan Project~\cite{HisExp_81}. The material was reflected by either BeO, Fe, WC, Pb, or Tu. Bare measurements (without a reflector) were also performed. 

\begin{figure*}[htb!]
	\centering\includegraphics{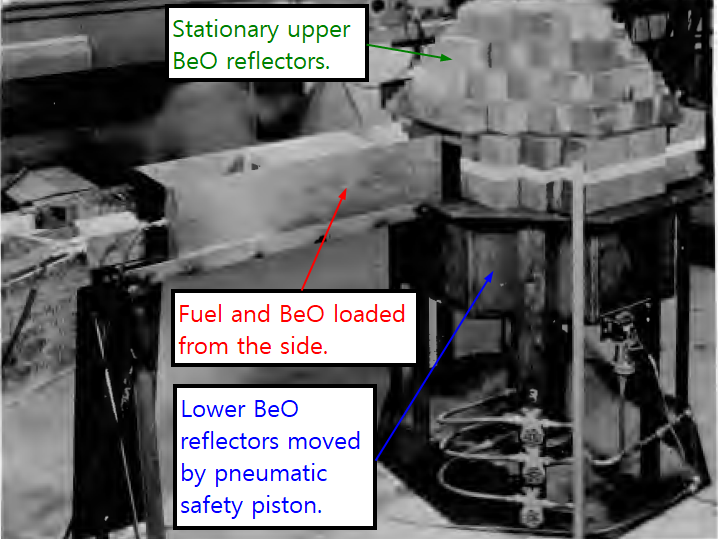}
	\caption{Hydride experiment assembly for BeO experiments~\cite{HisExp_80}.} 
	\label{Fig_F6_hydrideBeO}
\end{figure*}

There were three types of assemblies used in the hydride experiment. The first was the BeO experiment, shown in Figure~\ref{Fig_F6_hydrideBeO}, which consisted of BeO bricks stacked on a stationary reflector in a pseudo-sphere. The lower BeO bricks were on a movable platform. The fuel (and part of the BeO reflector) were in a steel tray that was inserted from the side. While other experiments (such as CP-1 and the Water Boiler) preceded this experiment, this is the first assembly that bears resemblance to the critical assembly machines used today. The operating process and safety functions are very similar to those used in today’s vertical lift assemblies (such as the Comet~\cite{HisExp_82} and Planet~\cite{HisExp_83} assemblies at NCERC). Some of the BeO experiments did not utilize this assembly and were built by hand with no safety features.

\begin{figure*}[htb!]
	\centering\includegraphics[width=4in]{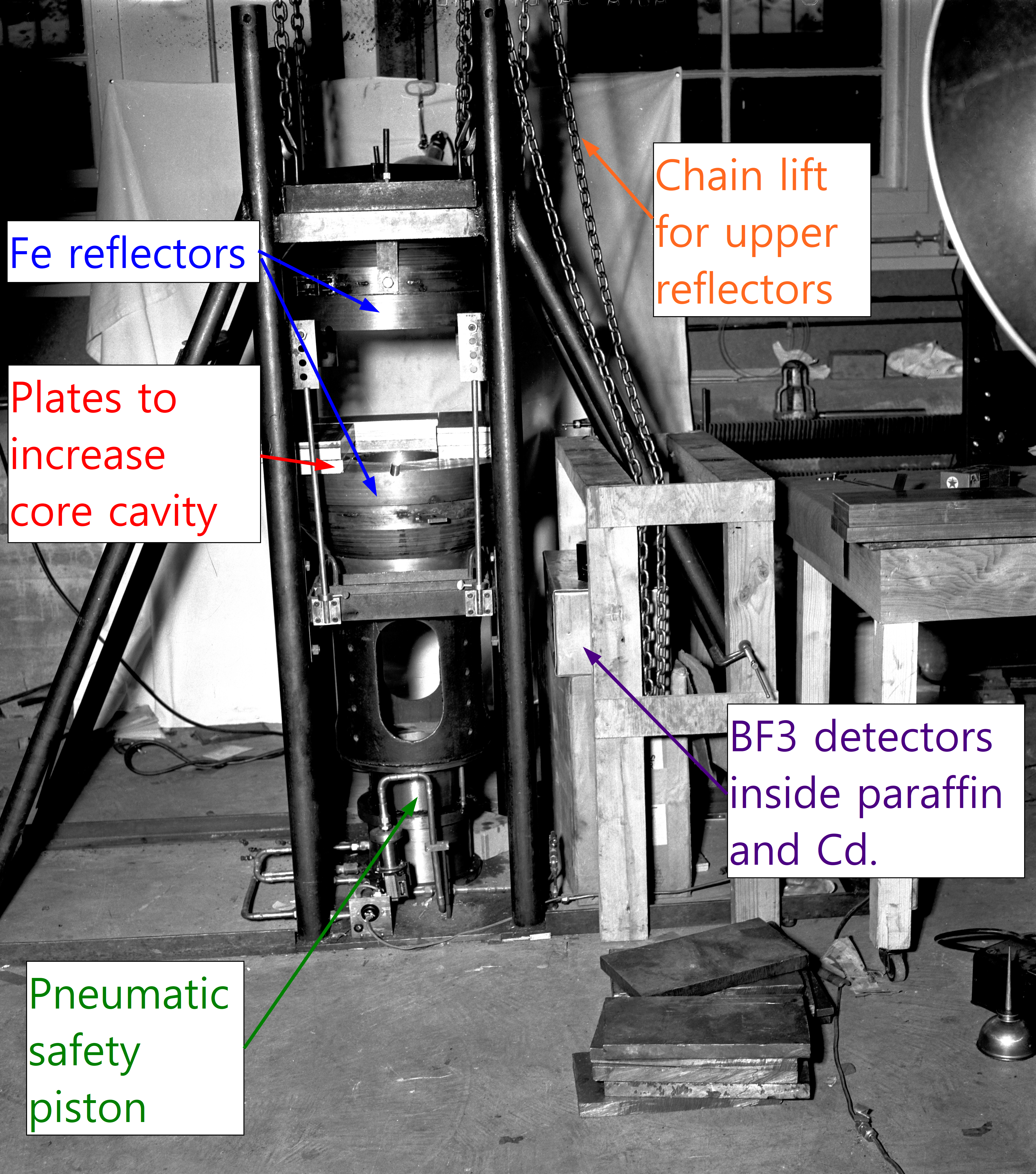}
	\caption{Hydride experiment assembly for the Fe, Tu, and Pb experiments (Fe reflector shown)~\cite{HisExp_80}.} 
	\label{Fig_F7_hydrideFe}
\end{figure*}

The second type of experiments used spherical Tu, Fe, and Pb reflection (with mating pieces to match the cube geometry). Unlike the BeO experiment, no material was inserted from the side. Here, the lower hemispherical reflector and hydride material were raised pneumatically into the upper hemispherical reflector. The Fe-reflected system is shown in Figure~\ref{Fig_F7_hydrideFe}. For the Tu, Fe, and Pb-reflected experiment, the system was not critical as a sphere. In order to obtain criticality, additional UH$_{10}$ cubes (and reflector blocks) were placed in between the two hemispheres, resulting in an elliptical system. When using the Pb reflector, not enough material was available, and criticality was never obtained.

Critical mass estimates were performed using a $1/M$ approach-to-critical, as shown in Table~\ref{Tab:3_hydride_PC}. All experiments were operated with personnel present in the room, as these experiments preceded the Daghlian and Slotin criticality accidents, which changed the way that critical experiments are performed~\cite{HisExp_84,HisExp_85,HisExp_86}. The experiments that utilized an assembly, however, would SCRAM if the neutron population in the external BF$_3$ detectors (shown in Figure~\ref{Fig_F7_hydrideFe}) exceeded a set threshold. Table~\ref{Tab:3_hydride_PC} includes comparison to simulations with MCNP6.2 with ENDF/B-VII.1 cross-sections.

\begin{table*}[bth!]
\caption{Hydride predicted critical results.}
\centering{}%
\begin{tabular}{|c|c|c|c|}
\hline
Reflector & Predicted critical (number of cubes) & Predicted critical mass (kg) & Simulated critical mass (kg)\tabularnewline
\hline
BeO & 920 & 5.91 $\pm$ 0.09	& 6.5\tabularnewline
\hline
WC & 1770 & 12.62 $\pm$ 0.2	& 12.8\tabularnewline
\hline
Tu & 1610 & 11.9 $\pm$ 0.2 & 10.6\tabularnewline
\hline
Fe & 1900 & 14.2 $\pm$ 0.3 & 14.6\tabularnewline
\hline
Pb & 2120 & 15.8 $\pm$ 0.3 & -\tabularnewline
\hline
Bare & 3240–4250 & 24.2–31.8 & 33.8\tabularnewline
\hline
\end{tabular}\label{Tab:3_hydride_PC}
\end{table*}

\begin{figure}[htb!]
	\centering\includegraphics[width=3in]{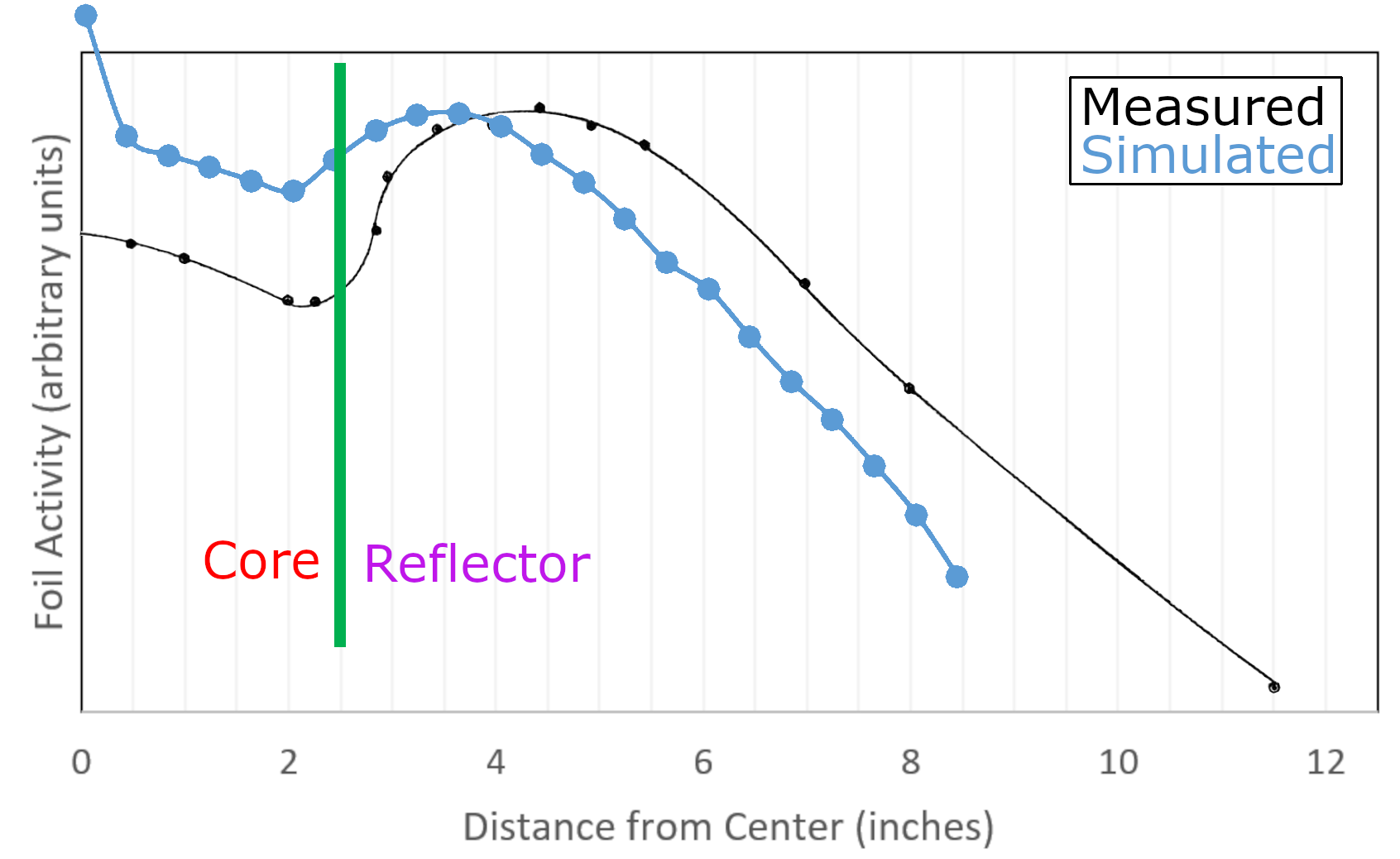}
	\caption{Au foil results for hydride experiment with BeO reflector.} 
	\label{Fig_F8_hydrideAu}
\end{figure}

\begin{figure}[htb!]
	\centering\includegraphics[width=3in]{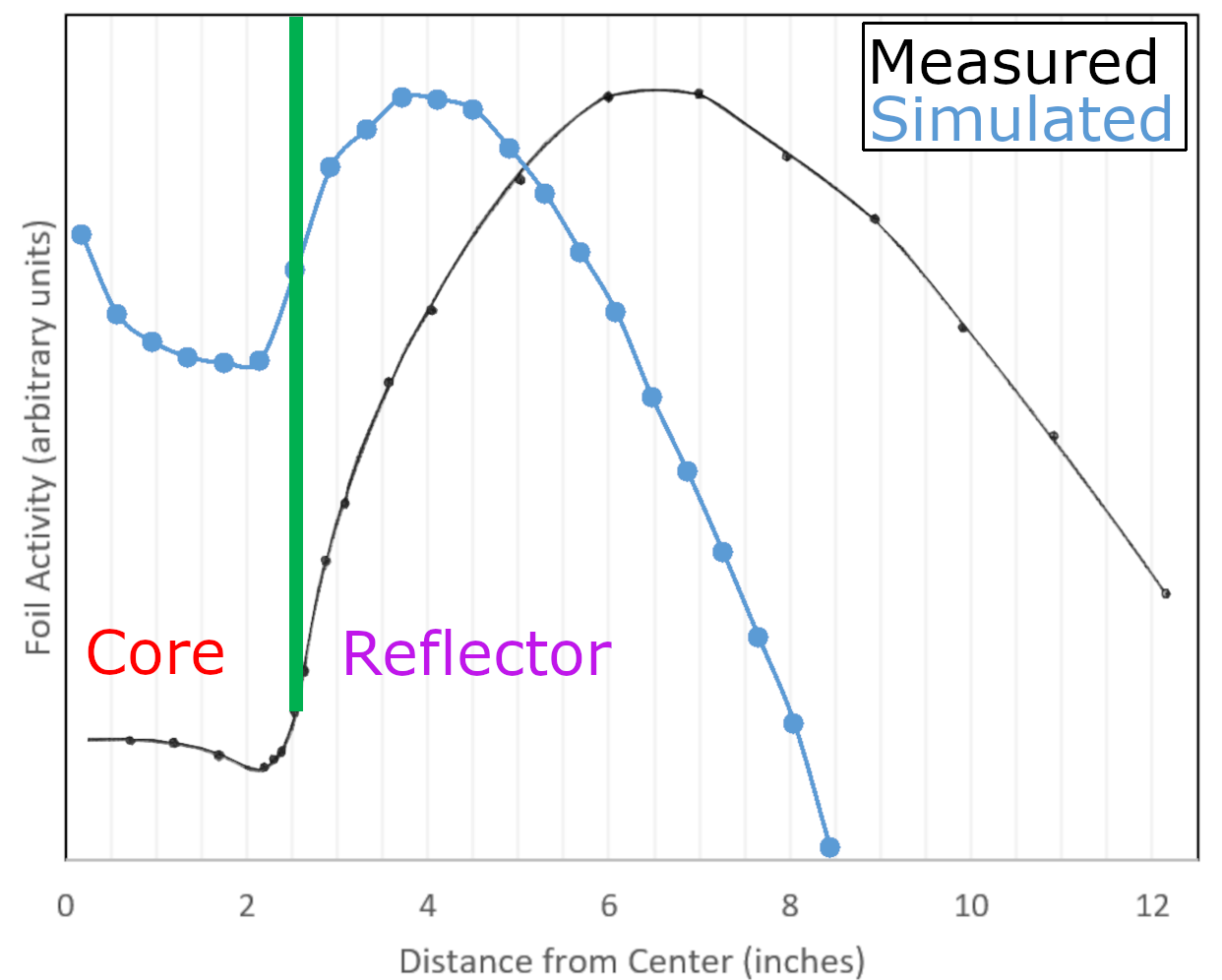}
	\caption{Mn foil results for hydride experiment with BeO reflector.} 
	\label{Fig_F9_hydrideMn}
\end{figure}

\begin{figure}[htb!]
	\centering\includegraphics[width=3in]{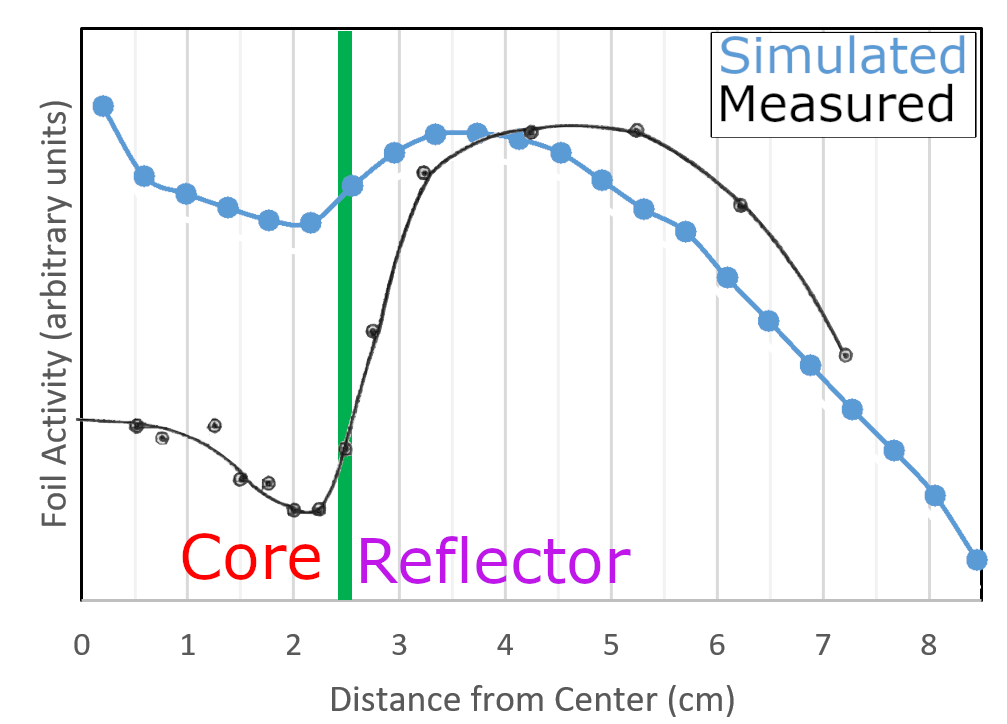}
	\caption{W foil results for hydride experiment with BeO reflector.} 
	\label{Fig_F10_hydrideW}
\end{figure}

Measurements were performed with Mn, Au, and W foils for the BeO- and WC-reflected experiments. These measurements provide spatial information of the neutron flux. A comparison of the measured and simulated (using MCNP6.2) foil activity is shown in Figures~\ref{Fig_F8_hydrideAu}-\ref{Fig_F10_hydrideW}. It can be seen that the Au foil results compare better than the Mn or W results. It is not clear if the poor comparison between measurement and simulations is due to large experiment uncertainties, a measurement bias, or a misunderstanding in the model geometry due to poor documentation. Similar measurements were performed by measuring the UH$_{10}$ parts themselves with a Geiger-Müller counter after going supercritical and ensuring that significant fissions had occurred to yield accurate statistics. These utilized UH$_{10}$ with dimensions of 0.125 x 0.5 x 0.5 inches or 0.25 x 0.5 x 0.5 inches; having smaller components allowed for greater spatial resolution. These experiments were also performed using BeO reflection with Cd in between the core and reflector regions. These results showed the importance of thermal neutron reflection from the BeO, and there was a large amount of thermal fission occurring at the edge of the core.

Rossi-$\alpha$ measurements~\cite{HisExp_87}, which provide the prompt neutron decay constant of the system, were also performed. These are neutron noise measurements that were first used during the Manhattan Project on the Water Boiler, and it was thus a very new measurement method when the hydride experiments were performed. This method utilizes knowledge that multiple neutrons are generally created at the exact same time from a single fission event. This resulted in neutron lifetime estimates of 5.5~$\mu$s and 1.3~$\mu$s for the BeO and WC experiments, respectively.

\section{Criticality accidents}
\label{sec:accidents}
The section will describe the world's first four criticality accidents, which occurred in Los Alamos in 1945 and 1946. Note that a criticality accident is any condition in which a supercritical state is achieved at a time when it is not intended. It is possible to have criticality accidents in which no personnel nor nuclear material nor equipment are harmed, although some criticality accidents result in loss of life (including some of the accidents described here). In many ways, the accidents described here had as much of an impact as the experiments themselves, as they established how criticality experiments should be safely performed. We still implement lessons learned from these accidents to this day.

\textbf{Accident 1}: The world's first criticality accident occurred with the Dragon~\cite{HisExp_6} assembly on February 11, 1945~\cite{HisExp_84}. Dragon used uranium hydride cubes, like those described in Section~\ref{sec:hydride}. A typical Dragon burst had a yield on the order of $10^{11}$ fissions. This could be modified by changing the starting neutron rate prior to the burst. If consecutive bursts were performed, the yield would continue to get larger. During the final burst, $6\cdot10^{15}$ fissions were produced. This resulted in the uranium hydride cubes rising in temperature so much that blistering and swelling occurred. The system expanded by about $1/8$ of an inch~\cite{HisExp_84,HisExp_1_40}. Given that no personnel were injured, and that the term "criticality accident" did not exist then, this was not considered a criticality accident at the time. The original report on Dragon does describe this final burst in detail, but does not refer to it as a criticality accident~\cite{HisExp_1_40}. It is noted that this incident does not necessarily meet the definition of a criticality accident given that the experimenters did intend to exceed superprompt critical. That being said, given that the yield was larger than expected and the fact that there was some noticeable material changes, this is commonly accepted today as a criticality accident and is included in the compilation of criticality accidents~\cite{HisExp_84}.

\textbf{Accident 2}: The second criticality accident occurred four months later, on June 6, 1945 (the one year anniversary of D Day). Here 35.4~kg of HEU metal (average 79.2\% 25) was built in a pseudosphere~\cite{HisExp_84,Accidents45to55}. This was in a polyethylene box which was put in a tank that was filled with water. The system went supercritical before the tank was fully filled. Subsequent inspection showed that the polyethylene box was not watertight. It was calculated that $3\textendash4\cdot10^{16}$ fissions occurred and the temperature in the metal may have rose by as much as 200$^\circ$C. There was no SCRAM system, and the configuration became subcritical due to the falling water level and the boiling of the water inside the polyethylene box (in addition to the reactivity loss of thermal expansion). There was non-lethal radiation exposure, and the material was used again for experiments three days later\cite{HisExp_84,Accidents45to55}. This event was also not considered to be a criticality accident at the time.

\begin{figure}[htb!]
	\centering\includegraphics[width=3in]{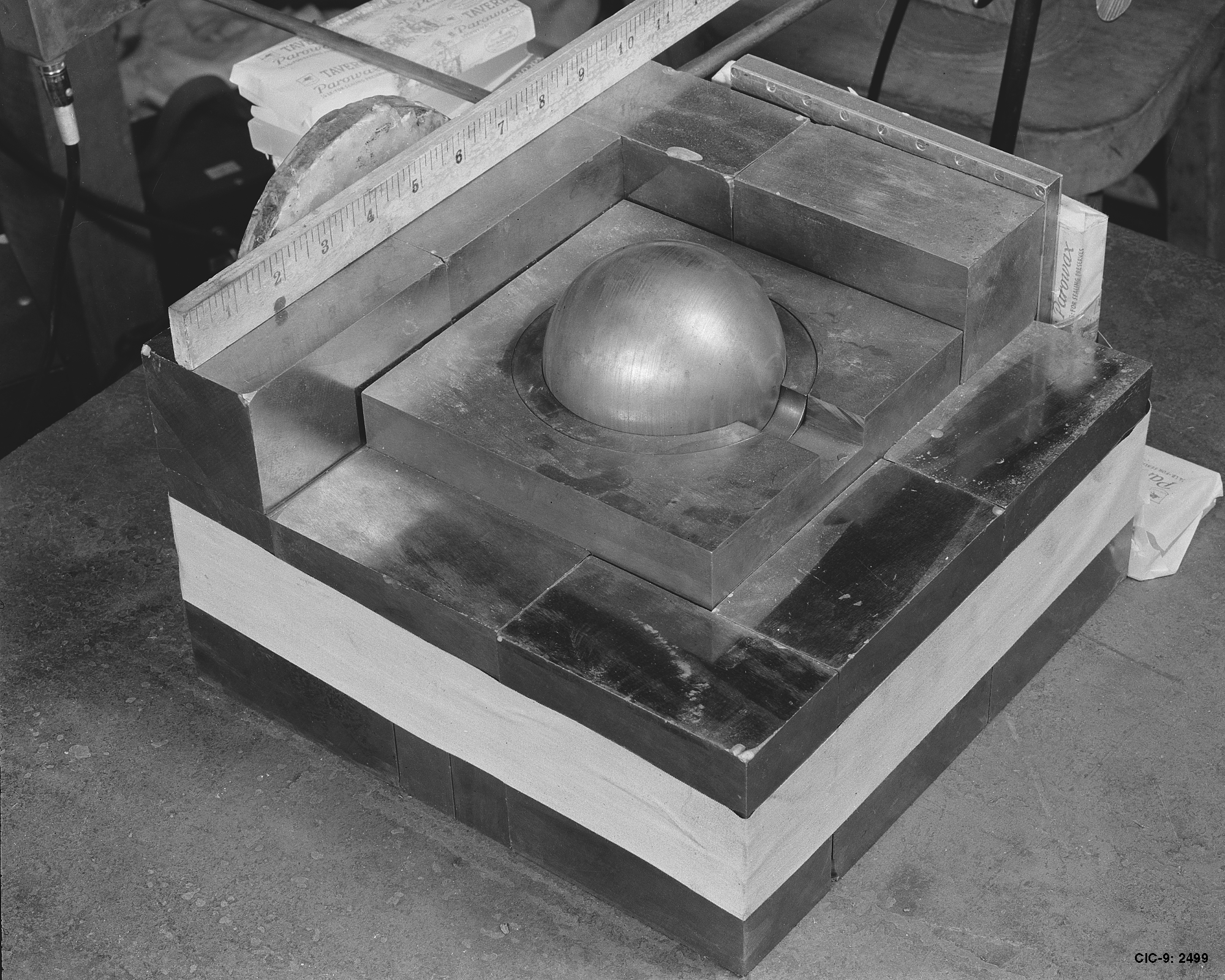}
	\caption{Pu sphere with WC reflection. This picture was taken as part of documentation related to the criticality accident on August 21, 1945 in which Harold Daghlian was killed~\cite{HisExp_84}.} 
	\label{Fig_F11_Daghlian}
\end{figure}

\textbf{Accident 3}: This accident involves a 6.2~kg sphere of $\delta$-phase Pu (with a density of 15.7~g/cm$^3$) reflected by WC~\cite{HisExp_84}. This can be considered an extension of the sphere measurements described in Section~\ref{sec:spheres}, which also included WC reflection. On August 21, 1945 Harold Daghlian was assembling this experiment (re-creation shown in Figure~\ref{Fig_F4_sphereM_ref}). He was working alone (although there was a guard 12 feet away)~\cite{Accidents45to55} late at night. He was going to add a WC brick to the assembly, but when he saw from the neutron counters that this would likely result in a supercritical system he withdrew his hand. The brick slipped out of his hand and fell onto the system, resulting in a superprompt critical configuration. He quickly pushed the brick off and unstacked the assembly. This accident resulted in $10^{16}$ fissions and a fatal dose of 510 rem (from which Harold Daghlian died 28 days later)\cite{HisExp_84}. While this was actually the third criticality accident, it was documented as the first accident at the time.

\textbf{Accident 4}: This accident involves the same core of Pu as the Daghlian accident. Louis Slotin (who became group leader of critical assemblies in late 1945)\cite{HisExp_1_49} was performing a demonstration. The configuration included the Pu core reflected by Be. In this accident, half of the upper Be hemisphere was touching the lower hemisphere while the edge 180$^\circ$ away was resting on a screwdriver. In this now famous criticality accident (which has been re-created in popular culture such as the 1989 "Fat Man and Little Boy" movie), the top hemisphere slipped, and the upper Be hemisphere contacted the lower hemisphere resulting in a supercritical configuration. Slotin quickly removed the upper hemisphere, but $3\cdot10^{15}$ fissions had already occurred. Slotin received over 1000~rem and died 9 days later\cite{HisExp_84,HisExp_1_50}. 

It should be noted that by today's standards, some might consider these four accidents "sloppy work" or think that the experimenters were not being careful or (worse yet) did not understand what they were doing. One must not, however, think about these isolated incidents without considering the larger context. It should be stressed that during 1945 thousands of troops were dying each day in the war, which resulted in immense time pressure. In addition, these experiments were essentially uncharted territory; these configurations are complex systems with 3D geometries and multiple materials, therefore criticality could not be estimated accurately at that time.

After the Slotin accident, the critical experiments capability was established at Pajarito Site (later called the Pajarito Laboratory and the Los Alamos Critical Experiments Facility). This established the protocol for conducting remote critical experiments\cite{HisExp_1_56,HisExp_1_57}. Some of the listed requirements include:
\begin{enumerate}
    \item Experiments will be conducted remotely unless it can be made critical only if several accidents, both independent and improbable occur simultaneously (known today as the "double contingency principle").
    \item The neutron population will be monitored and the speed of the assembly should be limited such that here is time for human response.
    \item The critical condition will be predicted from subcritical configurations (approach-to-critical methodology).
    \item Each assembly should have two independent safety devices that will disassemble the reactor sufficiently to stop the chain reaction.
    \item Experiments will be performed using written operating procedures.
\end{enumerate}

\section{Conclusions and continued impact}
\label{sec:conclusion}
Results of two experiments conducted during the Manhattan Project were discussed. The 25 and 49 experiments included the first subcritical experiments ever performed on kilogram quantities of metal nuclear material. They also provided the first multiplication results for bare and reflected 25 metal. The experiments described here helped establish criticality experiment operations. Critical experiments are extremely useful for nuclear data validation to this day; in fact, many of the early Los Alamos experiments from the 1950s (which grew out of these experiments) are some of the primary experiments used for validation of 25 and 49 nuclear data~\cite{HisExp_42}.  

These two experiments have continued impact on both Radiation Test Object (RTO) and critical assembly operations. The theories developed in conjunction with these experiments are used to this day. The approach-to-critical methodology used in these experiments (and previously at CP-1 and the Water Boiler) are still used today to estimate critical masses~\cite{HisExp_88}. Similarly, fission chamber and activation foils are used currently with critical experiments to provide nuclear data validation~\cite{HisExp_89}. The Rossi-$\alpha$ method is also still used to provide timing information on nuclear material systems~\cite{HisExp_90,HisExp_91}.

RTOs are configurations assembled primarily for radiation detection testing and validation. The 25 and 49 metal spherical experiments are essentially the first RTOs ever built. Today, hundreds of RTO operations are performed each year at NCERC~\cite{HisExp_92}. These RTOs also often involve spherical configurations with 25~\cite{HisExp_93,HisExp_94,HisExp_95} and 49~\cite{HisExp_96,HisExp_97,HisExp_98,HisExp_99,HisExp_100}. Modern experiments still have goals to provide measured results for improvements in nuclear data (such as the $\nu-1-\alpha$ results in Table~\ref{Tab:2_nu-1-alpha}). 

While the hydride work was ultimately stopped due to higher than expected critical masses and slow-neutron lifetime~\cite{HisExp_80}, this experiment included the first critical assembly machine that is similar to those used to this day. At NCERC there are four critical assembly machines that are routinely used for critical experiments, and two of them~\cite{HisExp_81,HisExp_82} are very similar to the hydride assemblies. While the hydride BeO assembly involves lifting reflector materials from the bottom and fuel with reflectors from the side, the current vertical lift machines generally only lift fuel and/or reflectors (and/or moderators) from the bottom (similar to the Tu, Fe, and Pb hydride experiments). The hydride experiments were performed with people in the same room, which would never be done today due to lessons learned from criticality accidents, but they did contain a SCRAM system (similar to those currently used). These experiments were not only important in the development of the Manhattan Project but have continued impact to this day.

The lessons learned from the four criticality accidents given in Section~\ref{sec:accidents} resulted in a protocol to perform critical experiments. These lessons learned were incorporated in the ANS-1 ("conduct of critical experiments") standard~\cite{ANS1}. The accidents described in this work are used for training of critical assembly operators to this day, at facilities such as NCERC. It should be noted that after this protocol described in Section~\ref{sec:accidents} was established, only one other experimental criticality accident resulted in fatalities has occurred in the US. Lessons learned from these early experiments and criticality accidents were also passed on to the metallurgists and chemists. These applied to the handling and fabrication of nuclear material and formed the basis of the field now known as criticality safety. These lessons learned are applied in nuclear facilities across the world.

\section*{Acknowledgments}
\label{sec:acknowledgements}
This work was supported by the US Department of Energy through the Los Alamos National Laboratory. Los Alamos National Laboratory is operated by Triad National Security, LLC, for the National Nuclear Security Administration of the US Department of Energy under Contract No. 89233218CNA000001.

Thanks to Daniel Alcazar, Alan Carr, and Mark Chadwick for help in finding historical documents and Richard Malenfant for discussions on historical experiments.

\vspace{0.25in}
\noindent\rule{0.35\textwidth}{.4pt}

\bibliographystyle{style/ans_js}  
  \small\bibliography{lacef-references-bold.bib}  


\end{document}